\documentclass[prl,twocolumn]{revtex4}
\usepackage{amssymb} 
\usepackage{amsmath} 
\usepackage{graphicx,epsfig}
\usepackage{mathrsfs} 
\usepackage[hypertex,linkcolor=red]{hyperref}
\newcommand{\kk}{\rangle\!\rangle} 
\newcommand{\bb}{\langle\!\langle}
\def\>{\rangle} \def\<{\langle}
\newcommand{\Tr}{\textrm{Tr}}
\def\Span{\operatorname{Span}}
\def\dim{\operatorname{dim}}
\def\Supp{\operatorname{Supp}} 
\def\Ker{\operatorname{Ker}}

 \def\N#1{|\!|{#1}|\!|} \def\sH{\mathcal{H}}
\def\sK{\mathcal{K}}
\renewcommand{\geq}{\geqslant}\renewcommand{\leq}{\leqslant}
\begin{document}
\title{Optimal data processing for quantum measurements}

\author{G. M. D'Ariano} 

\affiliation{QUIT Group,
  Dipartimento di Fisica ``A. Volta'', via Bassi 6, I-27100 Pavia,
  Italy and CNISM.} 

\author{P. Perinotti} 

\affiliation{QUIT Group,
  Dipartimento di Fisica ``A. Volta'', via Bassi 6, I-27100 Pavia,
  Italy and CNISM.} 

\begin{abstract}
  We consider the general measurement scenario in which the ensemble
  average of an operator is determined via suitable data-processing of
  the outcomes of a quantum measurement described by a POVM. We
  determine the optimal processing that minimizes the statistical
  error of the estimation.
\end{abstract}

\maketitle 

A measurement in Quantum Mechanics is usually associated to an {\em
  observable} represented by a selfadjoint operator $X$ on the Hilbert
space $\sH$ of the quantum system \cite{vonn}, with the eigenvalues
$x_i$ defining the possible outcomes of the measurement. The
probability distribution of the $i$th outcome is given by the Born
rule
\begin{equation}\label{Born}
p(i|\rho)=\Tr[P_i\rho]
\end{equation}
$\rho$ being the density operator of the state and $P_i$ denoting the
orthogonal projectors in the spectral decomposition $X=\sum_{i=1}^N
x_iP_i$ (for the sake of illustration here we consider only finite
spectrum). Consequently, the expected value for the outcome-averaging
over repeated measurements is given by the ensemble average
$\<X\>=\Tr[\rho X]$, with statistical error proportional to the r.m.s.
$\sqrt{\<\Delta X^2\>}$, with $\Delta X^2:=X^2-\<X\>^2$.

There are, however, more general kinds of measurements that can be
performed in the lab, which are not necessarily associated to any
observable, nevertheless enable the experimental determination of
ensemble averages: these are the measurements that are described by
POVM's. A POVM (acronym for Positive Operator-Valued Measure) is a set
of (generally nonorthogonal) positive operators $P_i\geq0$, $1\leq
i\leq N$ which resolve the identity $\sum_{i=1}^NP_i=I$ similarly to
the orthogonal projectors of an observable, whence with the same Born
rule (\ref{Born}). This more general class of quantum measurements
includes also the description of optimal joint measurements of
non-commuting observables \cite{jointmeas1,jointmeas2}, along with the
measurements of parameters with no corresponding observable such as
the phase of a harmonic oscillator \cite{phase}, and many other
practical measurements such as optimized discrimination of states for
quantum communications \cite{chefles}, and, most interesting, the so-called {\em
  informationally complete measurements} \cite{infocandgrouprep},
i.~e.  measurements that allow to determine the density matrix of the
state or any other desired ensemble average, as for the so-called
Quantum Tomography \cite{tomo}. Moreover POVM's also allow to provide
a full description of the measurement apparatus, including noisy
channels before detection \cite{clean}.  The POVM's are not just a
theoretical tool, since there is a general {\em quantum calibration}
procedure in order to determine experimentally the POVM of a
measurement device by using a reliable standard \cite{tomodev}.

How can we experimentally determine the ensemble average of the
(generally complex) operator $X$ using a POVM? Clearly this is
possible if $X$ can be expanded over the POVM elements (mathematically
we denote this condition as $X\in\Span\{P_i\}_{i=1,N}$. This means
that there exists a set of coefficients $f_i[X]$ such that
\begin{equation}
X=\sum_{i=1}^Nf_i[X]P_i,\quad X\in{\mathcal S}:=\Span\{P_i\}_{i=1,N}\label{expanX}
\end{equation}
When ${\mathcal S}\equiv\mathcal{B}(\sH)$ (i.~e. when all operators
can be expanded over the POVM), then the measurement is
informationally complete.  Obviously, once the expansion
(\ref{expanX}) is established one can obtain the ensemble average of
$X$ by the following averaging
\begin{equation}
\<X\>=\sum_{i=1}^Nf_i[X]p(i|\rho),
\end{equation}
where the probability distribution is given in Eq. (\ref{Born}).

The above general measurement procedure opens the problem of finding the coefficients $f_i[X]$ in
Eq. (\ref{expanX}), namely the data-processing of the measurement outcomes needed to determine the
ensemble average of $X$. In general the coefficients $f_i[X]$ are not unique (if
$N>\dim(\mathcal{S})$), and one then wants to optimize the data-processing according to a practical
criterion, typically minimizing the statistical error. This problem has never been addressed in the
general case, and its solution will be presented in this Letter. Notice that although the processing
functions are intrinsically linear in the definition (\ref{expanX}), there is no guarantee that the
optimal ones are linear in $X$. However, as we will see, remarkably the optimal processing function
is indeed linear in $X$, and depends only on the POVM and, in a {\em Bayesian scheme}, on the
ensemble of possible input states (due to the simplicity and popularity of the Bayesian scheme, in
this letter we will restrict the analysis only to this scheme, postponing the analysis of the {\em
  minimax} strategy to another more technical publication: for a comparison between the two
frameworks see, for example, Ref. \cite{Kahn}).  The derivation of the optimal data-processing
function requires some notions of frame theory \cite{fram,banfram} and linear algebra, which will be
introduced in the first part of the letter. Actually, for simplicity, instead of presenting the
actual derivation we will first prove uniqueness of the optimal processing, then we present the
result and prove that it satisfies the equations for optimality.  At the end we will also consider a
simple example of application for the sake of a quantitative estimation, showing that the
optimization can lead to sensible improvements.

In the Bayesian scheme one has an {\em a priori} ensemble ${\mathcal
  E}:=\{\rho_i, p_i\}$ of possible states $\rho_i$ of the quantum
system occurring with probability $p_i$.

For finite dimension all bounded operators are Hilbert Schmidt, whence
${\mathcal S}$ is a Hilbert space, and indeed ${\mathcal
  S}\subseteq\sH^{\otimes 2}$ and linear operators can be associated
to bipartite vectors as follows \cite{bellobs}
\begin{equation}
  A=\sum_{m,n=1}^dA_{mn}|m\>\<n|\leftrightarrow |A\kk=\sum_{m,n=1}^dA_{mn}|m\>|n\>,
\label{corresp}
\end{equation}
with the Hilbert-Schmidt scalar product $\bb A|B\kk\doteq\Tr[A^\dag B]$.
In the following we will retain the double-ket notation as a remind of
the correspondence (\ref{corresp}).  Completeness of the set of
vectors $\{|P_i\kk\}_{1\leq i\leq N}$ with ${\mathcal
  S}:=\Span\{|P_i\kk\}_{1\leq i\leq N}$ can be written as follows
\begin{equation}\label{framedef}
  a\N{X}^2_2\leq\sum_{i=1}^N|\bb P_i|X\kk|^2\leq b\N{X}_2^2,\quad X\in{\mathcal S}.
\end{equation}
with $0<a\leq b<\infty$, and the norm $\N{Z}_2$ is the Hilbert-Schmidt norm induced by the scalar
product $\N{Z}_2=\sqrt{\bb Z|Z\kk}=\sqrt{\Tr[Z^\dag Z]}$.  In the literature Eq.  (\ref{framedef})
with $|P_i\kk$ regarded as abstract vectors in the linear space ${\mathcal S}$ \cite{banach} define a so-called
{\em frame} of vectors. The main theorem of frame theory states that a set of vectors
in $\mathcal{S}$ is a frame iff the operator
\begin{equation}
  F=\sum_i|P_i\kk\bb P_i|,
\end{equation}
called {\em frame operator} is invertible \cite{fram} (here the fact that the set
$\{|P_i\kk\}_{1\leq i\leq N}$ is a frame for ${\mathcal S}$ trivially follows from the definition of
${\mathcal S}$. Since $F$ is invertible, one can obtain suitable coefficients $f_i[X]$ for the
expansion of a vector $|X\kk$ by the formula
\begin{equation}\label{canon}
  f_i[X]=\bb \Delta_i|X\kk
\end{equation}
where $\{\Delta_i\}$ is the {\em canonical dual} \cite{fram}, which is
defined through the identity
\begin{equation}
  |\Delta_i\kk=F^{-1}|P_i\kk.
\end{equation}
However, if the vectors $\{|P_i\kk\}_{1\leq i\leq N}$ are linearly
dependent, the processing rule (\ref{canon}) is not unique, and all
different choices of coefficients are provided by $f_i[X]=\bb
D_i|X\kk$, with $\{D_i\}$ are {\em alternate duals}. All alternate
duals can be classified as follows \cite{li}
\begin{equation}
  |D_i\kk=|\Delta_i\kk+|Y_i\kk-\sum_j |Y_j\kk\bb P_j|\Delta_i\kk,
\end{equation}
where the operators $\{Y_i\}$ are arbitrary elements of ${\mathcal
  S}$. Now, one can define a linear map $\Lambda$ from an abstract
$N$-dimensional space $\sK$ of coefficient vectors $|c\>$ to
$\mathcal{S}$ as follows
\begin{equation}
  \Lambda |c\>=\sum_{i=1}^Nc_i|P_i\kk,
\end{equation}
and $\Lambda$ has matrix elements $\Lambda_{mn,i}=(P_i)_{mn}$. By
definition any alternate dual must satisfy
\begin{equation}
  \sum_{i,j=1}^N\sum_{m,n=1}^d
  (P_j)_{pq}(D_j^*)_{mn}(P_i)_{mn}c_i=\sum_{i=1}^N(P_i)_{pq}c_i,
\end{equation}
for all $|c\>\in\sK$. Defining the matrix $\Gamma$ with elements
$(\Gamma)_{i,mn}=(D^*_i)_{mn}$ one has
\begin{equation}
  \Lambda\Gamma\Lambda=\Lambda,
\end{equation}
which is the definition of generalized inverse (or pseudoinverse) of
$\Lambda$. Alternate duals are then in one-to-one correspondence with
generalized inverses of $\Lambda$. This fact was already noticed in
Ref.  \cite{infolocvsglob}, and will be very useful in the
following.\par

We want now to minimize the statistical error in the determination of
the ensemble average. This is provided by the variance
\begin{equation}
  \delta_D(X):=\sum_{j=1}^Np(j|\rho_\mathcal{E})|f_j[X]|^2-\overline{|\<X\>|^2}_{\mathcal E},
\end{equation}
where $\rho_{\mathcal E}=\sum_ip_i\rho_i$, and
$\overline{|\<X\>|^2}_{\mathcal E} =\sum_ip_i|\Tr[\rho_iX]|^2$ is the
squared modulus of the expectation of $X$ averaged over the states in
the ensemble.  One has
\begin{equation}
  \delta_D(X)=\sum_{i=1}^N|\bb D_i|X\kk|^2\Tr[\rho_{\mathcal E} P_i]-\overline{|\<X\>|^2}_{\mathcal
    E}, 
\end{equation}
Notice that the term $\overline{|\<X\>|^2}_{\mathcal E}$ depends only
on the ensemble, and is independent of the POVM, whence we will focus
attention only on the contribution
\begin{equation}
  \Sigma_D (X)=\sum_{i=1}^N|\bb D_i|X\kk|^2\Tr[\rho_{\mathcal E} P_i].
\label{sigma}
\end{equation}
A relevant case is that of the {\em uniform ensemble}, with all pure
states equally distributed, corresponding to $\rho_{\mathcal E}=\frac
Id$ and $\overline{|\<X\>|^2}_{\mathcal E}=\frac1{d(d+1)}(\Tr[X^\dag
X]+|\Tr[X]|^2)$ \cite{infolocvsglob}.

Eq. (\ref{sigma}) defines a norm $\N{f_i[X]}^2_\pi$ of the vector of
coefficients corresponding to the metric matrix
$\pi_{ij}=\Tr[\rho_{\mathcal E} P_i]\delta_{ij}$.  Then, minimizing
$\Sigma_D (X)$ corresponds to determining the {\em minimum norm}
generalized inverse $\Gamma$ of $\Lambda$ with respect to the norm
$\N{c}_\pi=\sum_{i=1}^N|c_i|^2\pi_{ii}$. The minimum norm condition
for $\pi=I$ corresponds to the Moore-Penrose generalized inverse
$\Gamma$ \cite{bhapat}, satisfying the three conditions:
$\Gamma\Lambda\Gamma=\Gamma$, $\Gamma\Lambda=\Lambda^\dag\Gamma^\dag$
and $\Lambda\Gamma=\Gamma^\dag\Lambda^\dag$. The Moore-Penrose
generalized inverse of a matrix $Z$ (also denoted as $Z^\ddag$) turns
out to be simply the inverse of $Z$ on its support $\Supp(Z)$ (the
support $\Supp(Z)$ of $Z$ is the orthogonal complement of the kernel
$\Ker(Z)$ of $Z$), and acts as the null matrix on $\Ker(Z)$.

Following the same lines of derivation for the Moore-Penrose
generalized inverse one can show that the minimum norm generalized
inverse for a generic $\pi$ is independent of $X$, and is defined by
the condition \cite{infolocvsglob}
\begin{equation}
  \pi\Gamma\Lambda=\Lambda^\dag\Gamma^\dag\pi.
  \label{minnorm}
\end{equation}
The matrix $\Gamma\Lambda$ has matrix elements
$(\Gamma\Lambda)_{ij}=\bb D_i|P_j\kk$. Eq. (\ref{minnorm}) rewritten in terms of the optimal dual 
$\{\hat D_i\}$ becomes
\begin{equation}
\bb\rho_\mathcal{E}|P_i\kk\bb \hat D_i|P_j\kk=\bb P_i|\hat D_j\kk\bb P_j|\rho_\mathcal{E}\kk.
\end{equation}
Upon summing over the index $i$, and remembering that for any dual $\{D_i\}$ one has
$\sum_i|P_i\kk\bb D_i|=\Pi_\mathcal{S}$ where $\Pi_\mathcal{S}$ is the projection on $\mathcal S$,
one has $\bb\rho_\mathcal{E}|P_j\kk=\Tr[\hat D_j]\bb P_j|\rho_\mathcal{E}\kk$, consequently
$\Tr[\hat D_i]=1$. This implies that the optimal processing function for the identity operator is
$f_i[I]=1$, whence $\delta_{\hat D}(I)=0$, whereas, remarkably, $f_i[I]$ is generally non constant
for the canonical dual.

We will now prove that the solution of Eq.~\eqref{minnorm} is unique.
For not invertible $\pi$ we can restrict Eq. (\ref{minnorm}) to
$\Supp(\pi)$, and from now on we will denote the corresponding blocks
of all matrices with the same symbols. Suppose now that there exist
two generalized inverses $\Gamma$ and $\Gamma'$ satisfying
Eq.~\eqref{minnorm}. Upon defining $\Theta=\Gamma-\Gamma'$, we have
that
\begin{equation}
  \left\{
\begin{array}{l}
  \Lambda\Theta\Lambda=0\\
  \pi\Theta\Lambda=\Lambda^\dag\Theta^\dag\pi,
\end{array}
\right.
\end{equation}
and multiplying on the left by $\Lambda\pi^{-1}$ both members of the
second equation, and substituting the first equation we obtain
$\Lambda\Theta\Lambda=\Lambda\pi^{-1}\Lambda^\dag\Theta^\dag\pi=0$, or
equivalently, by invertibility of $\pi$,
$\Lambda\pi^{-1}\Lambda^\dag\Theta^\dag=0$. The matrix
$\Lambda\pi^{-1}\Lambda^\dag$ can be rewritten as
\begin{equation}
  \Lambda\pi^{-1}\Lambda^\dag=\sum_{i=1}^N(\pi_{ii})^{-1}|P_i\kk\bb P_i|.
\end{equation}
Since $\Lambda\pi^{-1}\Lambda^\dag\geq0$, a sufficient condition for a
vector $X\in\mathcal{S}$ to be in $\Ker(\Lambda\pi^{-1}\Lambda^\dag)$
is that $\bb X|\Lambda\pi^{-1}\Lambda^\dag|X\kk=0$, namely
\begin{equation}
  \sum_{i=1}^N(\pi_{ii})^{-1}|\bb X|P_i\kk|^2=0,\;\Leftrightarrow\;X\in\mathcal{S},
\end{equation}
which is possible iff $\bb X|P_i\kk=0$ for all $i$. By completeness of
$P_i$, this is equivalent to say that the only vector of $\mathcal{S}$
in $\Ker(\Lambda\pi^{-1}\Lambda)$ is $X=0$. Then
$\Lambda\pi^{-1}\Lambda$ is full rank, whence $\Theta=0$, or
equivalently $\Gamma=\Gamma'$.\par

We will now provide the solution to Eq.~\eqref{minnorm} in terms of
the optimal dual, which is expressed as
\begin{equation}
  \hat D_i=\Delta_i-\sum_{j=1}^N([(I-M)\pi(I-M)]^\ddag\pi M)_{ij}\Delta_j,
\label{optdual}
\end{equation}
where $\Delta_i$ is the canonical dual, $M_{ij}=\bb\Delta_i|P_j\kk=\bb
P_i|F^{-1}|P_j\kk=\bb P_i|\Delta_j\kk=M_{ji}^*$.  Since
$\Delta_i^\dag=\Delta_i$, $M_{ij}^*=M_{ij}$ \cite{selfadj} and the
optimal dual frame $\{\hat D_i\}$ in Eq.~\eqref{optdual} is
selfadjoint because the matrix $[(I-M)\pi(I-M)]^\ddag\pi M$ has real
elements. Notice that $M^2=M$ and $M^\dag=M$, namely $M$ is an
orthogonal projector, as can be easily verified. Also $(I-M)$ is an
orthogonal projector, and
$[(I-M)\pi(I-M)]^\ddag(I-M)=[(I-M)\pi(I-M)]^\ddag$. The matrix
$\Gamma\Lambda$ for the optimal dual frame can be easily calculated,
and is equal to
\begin{equation}
  \Gamma\Lambda=M-[(I-M)\pi(I-M)]^\ddag\pi M.
\end{equation}
We can substitute this expression in Eq.~\eqref{minnorm} to verify its
validity. We have indeed
\begin{equation}
\begin{split}
  &\pi\Gamma\Lambda=\pi M-\pi[(I-M)\pi(I-M)]^\ddag\pi M\\
  =&\pi M+\pi[(I-M)\pi(I-M)]^\ddag(I- M)\pi(I- M)\\
  &-\pi[(I-M)\pi(I-M)]^\ddag\pi\\
  =&\pi-\pi[(I-M)\pi(I-M)]^\ddag\pi,
\end{split}
\end{equation}
and analogously
\begin{equation}
  \Lambda^\dag\Gamma^\dag\pi=\pi-\pi[(I-M)\pi(I-M)]^\ddag\pi.
\end{equation}
When $\pi\propto I$ the canonical dual is optimal, since for the
canonical dual one has $\Gamma\Lambda=\Lambda^\dag\Gamma^\dag$. This
is the case e.g. of the uniform ensemble of pure states with POVM
elements with constant trace, which includes all {\em covariant} POVMs
studied in Ref.  \cite{infolocvsglob}. In the general case, one can
write the expression of Eq.~\eqref{sigma} as follows
\begin{equation}
  \Sigma_{\hat D}(X)=\Sigma_\Delta(X)-\Psi(X),
\end{equation}
where $\Sigma_\Delta$ is the contribution of the canonical dual
\begin{equation}
  \Sigma_\Delta(X)=\sum_{i=1}^N|\bb \Delta_i|X\kk|^2\Tr[\rho_{\mathcal E} P_i],
\end{equation}
and $\Psi$ is the correction due to the optimization which is given by
\begin{equation}
  \Psi(X)=\sum_{i,j=1}^N\bb X|\Delta_i\kk(\pi[(I-M)\pi(I-M)]^\ddag\pi)_{ij}\bb\Delta_j|X\kk.
  \label{epsilon}
\end{equation}
The relative added noise of the canonical dual compared to the optimal
one is given by
\begin{equation}
  \epsilon(X)\doteq\frac{\delta_\Delta(X)-\delta_{\hat D}(X)}{\delta_{\hat D(X)}}=\frac{\Psi(X)}{\Sigma_\Delta(X)-\Psi(X)-\overline{|\<X\>|^2}_{\mathcal E}}.
\end{equation}
\begin{figure}[h]
\psfig{file=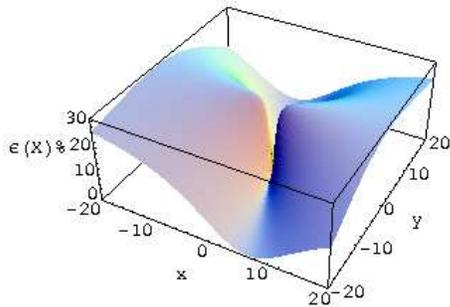,width=6cm}
\caption{\label{valle} Example of optimized data-processing rule for the informationally complete
  POVM in Eq.~\eqref{povm}. The plot shows the relative added noise in Eq.~\eqref{epsilon} for
  $X=\sigma_z+x\sigma_x+y\sigma_y$ versus $x$ and $y$}
\end{figure}
A quantitative estimate of $\epsilon(X)$ can be obtained from the
following example in dimension two (see Fig. \ref{valle}). Consider
the POVM
\begin{equation}
\begin{split}
  P_1&=\begin{pmatrix}
    \frac{64}{1197}&-\frac{16}{1197}\\
    -\frac{16}{1197}&\frac{40}{1197}
  \end{pmatrix},\ P_2=\begin{pmatrix}
    \frac{34}{1197}&\frac{2(1-16i)}{1197}\\
    \frac{2(1+16i)}{1197}&\frac{34}{1197}
  \end{pmatrix},\\
  P_3&=\begin{pmatrix}
    \frac{281}{399}&-\frac{18-32i}{1197}\\
    -\frac{18+32i}{1197}&\frac{289}{399}
  \end{pmatrix},\ P_4=\begin{pmatrix}
    \frac{64}{399}&\frac{64(1-i)}{1197}\\
    \frac{64(1+i)}{1197}&\frac{32}{399}
  \end{pmatrix},\\
  &P_5=\begin{pmatrix}
    \frac{64}{1197}&\frac{-32(1+2i)}{1197}\\
    -\frac{32(1-2i)}{1197}&\frac{160}{1197}
  \end{pmatrix}.
\end{split}
\label{povm}
\end{equation}
The operator $X$ is the following selfadjoint operator
\begin{equation}
X=
\begin{pmatrix}
1&-2.24 +i\\
-2.24 - i&-1
\end{pmatrix},
\end{equation}
and for an ensemble of uniformly distributed pure states
$\frac16(\Tr[X^2]+\Tr[X]^2)=\frac16\Tr[X^2]=2.34$. By direct
calculation one obtains $\Sigma_\Delta=799.66$ and $\Psi=133.05$, and
finally
\begin{equation}
\epsilon(X)\simeq0.2,
\end{equation}
which means a relative added noise of about $20\%$. This example shows
that a correct processing can highly improve the statistics of
expectation values, and eventually the convergence rate of tomographic
state reconstruction. The additional error due to the use of the
canonical dual instead of the optimal one is equivalent to a
depolarizing channel with depolarization probability $0.09$.\par

In conclusion, we considered the general measurement scenario in which
the ensemble average of an operator is determined via suitable
data-processing of the outcomes of a quantum measurement described by
a POVM. We have determined the optimal processing that minimizes the
statistical error of the estimation. Contrarily to the widespread
conviction, the optimal data-processing is generally not obtained via
the canonical dual of the POVM, and the improvement due to
optimization can be substantial. The present analysis has been carried
out for finite spectrum and finite dimensions, however, it can be
easily generalized to discrete spectrum in infinite dimensions for
bounded operators and bounded duals, and, with more technicalities,
even to continuous spectrum (the case of quantum homodyne tomography
\cite{tomo}). We believe that the present result will allow to improve
greatly many relevant experimental analysis of quantum measurements.

\acknowledgments P. P. thanks L. Maccone and M. F. Sacchi for interesting discussions and
suggestions. This work has been supported by Ministero Italiano dell'Universit\`a e della Ricerca
(MIUR) through PRIN 2005. P. P.  acknowledges financial support by EC under pro ject SECOQC
(contract n. IST-2003-506813)


\begin{thebibliography}{99}
\bibitem{vonn} J. Von Neumann, {\em Mathematical Principles of Quantum
    Mechanics}, (Princeton University Press, Princeton, 1955).
\bibitem{jointmeas1} E. Arthurs and J. L. Kelly, Bell. Syst. Tech. J., {\bf 44} 725-729 (1965).
\bibitem{jointmeas2}  J. P. Gordon and W. H. Louisell, in {\em Physics of Quantum Electronics}, pp. 833-840,
  McGraw-Hill, (New York, 1966). 
\bibitem{phase} C. W. Helstrom {\em Quantum Detection and Estimation
    Theory},(Academic Press, New York, 1976).
\bibitem{chefles} A. Chefles, Phys. Rev. A {\bf 64} 062305 (2001).
\bibitem{infocandgrouprep} G. M. D'Ariano, P. Perinotti, and M. F.  Sacchi, Europhys. Lett. {\bf
    65}, 165 (2004); J. Opt. B: Quantum Semiclass. Opt. {\bf 6}, S487 (2004); 
\bibitem{tomo} G. M. D'Ariano, M. G. A. Paris, and M. F. Sacchi, Adv.
  Imaging Electron Phys. {\bf 128}, 205 (2003).
\bibitem{clean} F. Buscemi, G. M. D'Ariano, M. Keyl, P. Perinotti, and
  R. F. Werner, J. Math. Phys. {\bf 46}, 082109 (2005).
\bibitem{tomodev} G. M. D'Ariano, P. Lopresti, and L. Maccone, Phys.
  Rev. Lett. {\bf 93}, 250407 (2004).
\bibitem{Kahn} G. M. D'Ariano, J. Kahn, and M. F. Sacchi, Phys. Rev. A
  {\bf 72}, 032310 (2005).
\bibitem{fram} R. J. Duffin and A. C. Schaeffer, Trans. Am. Math.
  Soc. {\bf 72}, 341 (1952); P. G. Casazza, Taiw. J. Math. {\bf 4},
  129 (2000).
\bibitem{banfram} P. Casazza, D. Han, and D. R. Larson, Contemp. Math.
  {\bf 247}, 149 (1999).
\bibitem{bellobs} G. M. D'Ariano, P. Lopresti, and M. F. Sacchi, Phys.
  Lett. A {\bf 272}, 32 (2000).
\bibitem{banach} In infinite dimension generally one considers
  ${\mathcal S}$ as a Banach space (see Ref. \cite{banfram}) and the
  condition $b<\infty$ is generally non trivial as in the finite
  dimensional case.
\bibitem{li} S. Li, Numer. Funct. Anal. Optim. {\bf 16}, 1181 (1995).
\bibitem{infolocvsglob} G. M. D'Ariano, P. Perinotti, and M. F.
  Sacchi, Phys. Rev. A {\bf 72}, 042108 (2005).
\bibitem{bhapat} R. B. Bhapat, {\em Linear Algebra and Linear Models},
  (Springer-Verlag, New York, 2000).
\bibitem{selfadj} Notice that since the swap operator $E$ acts on a
  vector $|A\kk\in\mathcal{L}(\sH)$ as $E|A\kk=|A^T\kk$, where $A^T$
  is the transpose of $A$ on the basis $|n\>$ of Eq.~\eqref{corresp},
  by selfadjointness of $P_i$ one has $E|P_i^*\kk=|P_i\kk$, and
  $EF^*E=F$. Similarly $EF^{-1*}E=F^{-1}$, and then
$$E|\Delta^*_i\kk=EF^{-1*}|P_i^*\kk=F^{-1}E|P_i^*\kk=|\Delta_i\kk,$$
namely $\Delta_i^\dag=\Delta_i$. As a consequence
$\bb\Delta_i|P_j\kk=\Tr[\Delta_iP_j]\in{\mathbb R}$.
\end{thebibliography}
\end{document}